\documentclass[aps,prl,floatfix,twocolumn,footinbib]{revtex4-1}

\usepackage{graphicx,bm,amsmath,amssymb,natbib,url,epsfig,color,hyperref}

\usepackage{mathrsfs}
\usepackage{amsthm}
\usepackage{enumerate}
\usepackage{float}
\newcommand*{\rom}[1]{\expandafter\@slowromancap\romannumeral #1@}
\begin{document}
\title{Negative Excess Shot Noise by Anyon Braiding}

\author{Byeongmok Lee}
\affiliation{Department of Physics, Korea Advanced Institute of Science and Technology, Daejeon 34141, Korea}

\author{Cheolhee Han}
\affiliation{Department of Physics, Korea Advanced Institute of Science and Technology, Daejeon 34141, Korea}
 
\author{H.-S. Sim}\email[]{hssim@kaist.ac.kr}
\affiliation{Department of Physics, Korea Advanced Institute of Science and Technology, Daejeon 34141, Korea}

\date{\today}
%\affiliation{}

\begin{abstract}
Anyonic fractional charges $e^*$ have been detected by autocorrelation shot noise at a quantum point contact (QPC) between two fractional quantum Hall edges. We find that the autocorrelation noise can also show a fingerprint of Abelian anyonic fractional statistics.
We predict the noise of electrical tunneling current $I$ at the QPC of the fractional-charge detection setup, when anyons are dilutely injected, from an additional edge biased by a voltage, to the setup in equilibrium.
At large voltages, the nonequilibrium noise is {\it reduced} below the thermal equilibrium noise by the value $2 e^* I$.
This negative excess noise is opposite to the positive excess noise $2e^* I$ of the conventional fractional-charge detection and also to usual positive autocorrelation noises of electrical currents.
This is a signature of the Abelian fractional statistics, resulting from the effective braiding of an anyon thermally excited at the QPC around another anyon injected from the additional edge.
\end{abstract}
\maketitle  
 
 %Fractional charge and fractional statistics are the basic features of anyons. The fractional charges $e^*$ were detected by shot noise at a quantum point contact (QPC) between two fractional quantum Hall edges. 

Abelian anyons appear in fractional quantum Hall (FQH) systems of filling factor $\nu = 1 / (2n+1)$, $n = 1,2,\cdots$.
They obey the fractional exchange statistics~\cite{Leinaas,Arovas,Stern}.
Two anyons gain the phase $\pm \pi \nu$ 
when their positions are adiabatically exchanged, and $\pm 2 \pi \nu$ when one braids around the other. 
%Detecting the fractional statistics in experiments requires efforts.
Proposals~\cite{Chamon,Vishveshwara,Kim06,Safi,Campagnano12,Law,Feldman,Kane03,Rosenow,An,
Camino, Willett, Ofek, McClure, Halperin,Rosenow2,Grosfeld} for detecting the fractional statistics are based on interferometers or current-current cross-correlations. They involve quantities experimentally inaccessible or affected by unintended setup change or Coulomb interaction. It will be useful to find fractional-statistics effects experimentally feasible.

 Shot noise $S$, zero-frequency nonequilibrium fluctuation of electrical current $I$, has valuable information~\cite{Blanter}.
Its Poisson value $S = 2q I$ in the tunneling regime of a quantum point contact (QPC) was
used to detect the charge $q$ of current carriers~\cite{Reznikov}. The fractional charge $e^* = \nu e$ of anyons was measured~\cite{Kane94,Goldman,Picciotto,Saminadayar,Dolev,Reznikov2,Griffiths} from the ratio $S/I = 2 e^*$ at a QPC between FQH edges; $e$ %$e>0$ 
is the electron charge. The Poisson value originates from uncorrelated transfer of discrete charges.
Reduction or enhancement from the value signifies effects such as resonances, diffusive scattering, Cooper pairing, etc~\cite{Blanter}.

\begin{figure}[b] 
\centering
\includegraphics[width=\columnwidth]{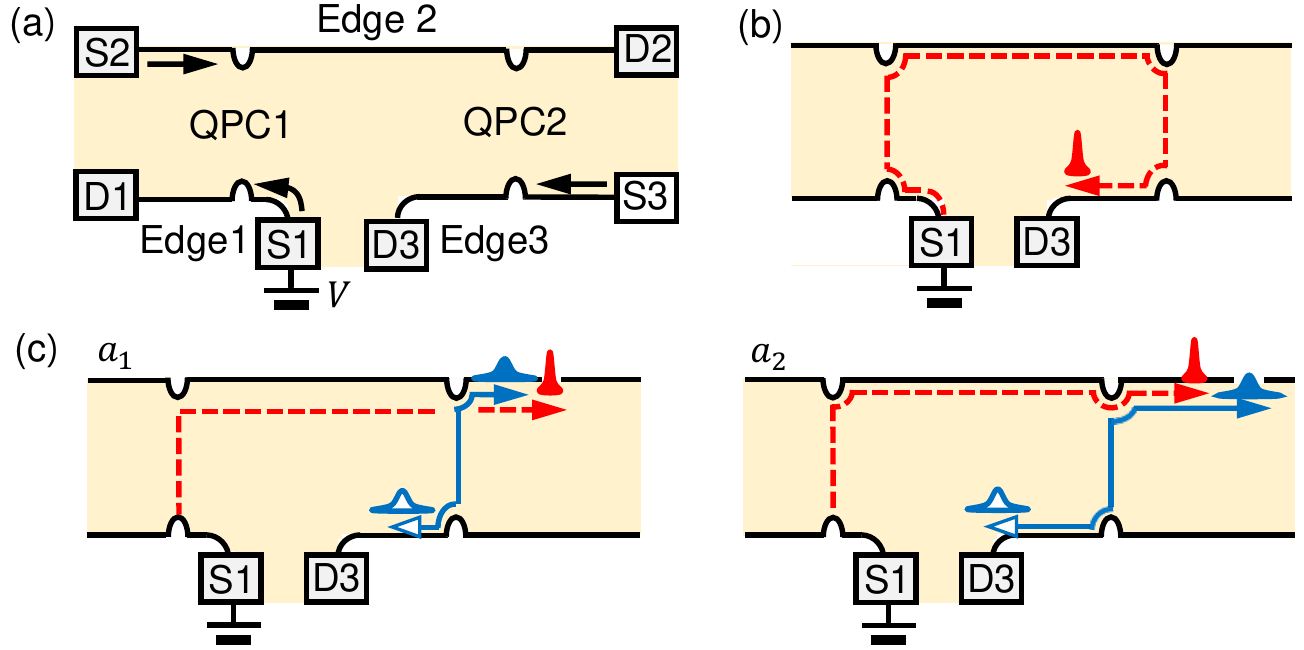}
% width=1\linewidth tells latex to resize the image to 1 linewidth. width=.5\linewidth would be half the linewidth. the stuff inside the {} is the path name to the image file. 
\caption{(a) Setup  at $\nu = \frac{1}{2n+1}$. Chiral edge channel Edge$i$ propagates (arrows) from source S$i$ to drain D$i$, $i=1,2,3$.  S1 is biased by voltage $V$, while the other sources and drains are grounded. Anyon tunneling occurs at QPC1 (QPC2) between Edge2 and Edge1 (Edge3). 
(b) Poisson process. 
A particle-like anyon 
%of size $\hbar v / (e^* V)$ 
biased by $V$ (narrow filled packets, dashed arrows) moves from Edge1 to D3 through tunneling at QPC1 and QPC2.
%, where $v$ is the anyon velocity. 
(c) Interference between subprocesses $a_1$ and $a_2$.
A particle-like anyon biased by $V$ moves (dashed) from Edge1 to D2 through tunneling at QPC1. After (before) this anyon passes QPC2 along Edge2, a particle-hole pair excitation thermally occurs at QPC2 in $a_1$ ($a_2$).
The particle-like anyon (wide filled packets) and the hole-like anyon (wide empty) in the pair
%, whose packet size is $\hbar v / k_B T$,
move (solid arrows) along Edge2 and Edge3, respectively.  
%When the overlap of the packets between $a_1$ and $a_2$ is nonvanishing, the process contributes to the thermal noise $S(0,T)$.  
The interference between $a_1$ and $a_2$ involves braiding of the thermal anyon around the voltage-biased anyon.
%, resulting in the negative excess noise.
}  \label{setup} 
\end{figure}

In this work, we predict unusual behavior of shot noise, originating from the Abelian fractional statistics of Laughlin anyons, in the setup [Fig.~\ref{setup}(a)] composed of the conventional fractional-charge detection part (Edge2, Edge3, QPC2) and an additional edge (Edge1).
%a simple modification of the conventional fractional-charge detection setup 
Anyons are dilutely injected~\cite{Comforti,Comforti2,Chung,Kane_PRB} via QPC1 from Edge1, biased by voltage $V$, to the detection part in equilibrium.
We find that the zero-frequency autocorrelation noise $S(V,T)$ of tunneling current $I$ at QPC2 is {\it reduced} below the thermal equilibrium noise $S(0,T)$ at temperature $T$,
\begin{equation}
\delta S = - 2 e^* I < 0 \quad \quad \quad \textrm{at} \quad e^*V \gg k_B T. \label{excessN}
\end{equation}
$\delta S \equiv S (V, T) - S(0, T)$ is the excess shot noise with respect to the thermal noise and $k_B$ is Boltzmann constant.
The negative excess noise is unusual, since the setup has the conventional Poisson process [Fig.~\ref{setup}(b)] enhancing the noise; it is opposite to the positive noise $2 e^* I > 0$ of the conventional fractional-charge detection~\cite{Kane94,Goldman,Picciotto,Saminadayar,Dolev,Reznikov2,Griffiths}. 
By contrast, in the integer quantum Hall regime at $\nu=1$, the setup shows the positive Poisson noise of $\delta S  = 2eI > 0$, which cannot be extrapolated from Eq.~\eqref{excessN} with $e^* =  e/(2n+1) \to e$.

The negative excess noise results from an interference involving anyon braiding [Fig.~\ref{setup}(c)], which weakens thermal anyon tunneling at QPC2, reducing the noise.
The reduction dominates over the enhancement by the Poisson process.
Interestingly, for electrons at $\nu=1$, the interference does not exist, as it is described by a pair of disconnected Feynman diagrams that exactly cancel each other, according to the linked cluster theorem~\cite{Fetter}.
For anyons, the cancellation is only partial, since the subdiagrams (vacuum bubbles) of one of the disconnected diagrams are linked~\cite{AlgebraicT} by the braiding. 
This type of anyon processes, vacuum bubbles linked by braiding, is called topological vacuum bubbles (TVBs)~\cite{Han}.
Detection of the negative excess noise is experimentally feasible, and will provide a signature of TVBs and the fractional statistics in the case of pristine edges (without edge reconstruction).
The signature manifests itself in the leading-order contributions (in QPC tunneling strengths) to the excess noise, thanks to the dilute anyon injection at QPC1.

%

%The present work shows that TVBs appear in the simpler setup of Fig.~\ref{setup}(a), and cause the unusual negative excess noise that is more easily detectable (with avoiding dephasing) than the interference current of Ref.~\cite{Han}.

{\it Excess noise.---} We consider the time $t$ average $I = \overline{I(t)}$ of tunneling current $I(t)$ at QPC2, and
its zero-frequency noise $S = 2 \int^\infty_{-\infty} dt (I(t) - I)(I(0) - I)$.
%{\color{red} symmetrization, factor two}
%\begin{equation}
%S = \int_{-\infty}^{\infty} dt \langle [  \hat{I}_{b}\left(0\right)-  I_{b}  ] [ \hat{I}_{b} \left(t\right)  - I_{b} ] \rangle.
%\end{equation} where $I_{b}= \langle\hat{I}_{b} \rangle$.
%The non-equilibrium excess noise of tunneling current and concomitant fano factor are defined as \begin{align}
%&\Delta S_{b,b} \equiv S_{b,b}\left(V,T \right)-S_{b,b}\left(0,T \right) \\ &\mathcal{F}\equiv \frac{\Delta S_{b,b}}{e^{*} I_{b} } \end{align} where $V$,$T$,$e^{*}$ denote bias voltage of Edge 1, system temperature, charge of a tunneling particle, respectively. 
Employing a perturbation theory based on the chiral Luttinger liquid~\cite{Wen,vonDelft}, Keldysh Green's functions, and Klein factors~\cite{Guyon}, we derive $I$ and $\delta S = S (V, T) - S(0, T)$ at voltages $e^* V \gg k_B T$ in the anyon tunneling regime of $\gamma_i T^{\nu-1} \ll 1$, up to the leading order $O (\gamma_1^2 \gamma_2^2)$ of tunneling strength $\gamma_i$ at QPC$i$, % (see Appendix for the derivation and $f(\nu)$),
\begin{align}
 I & \simeq e^* \gamma_1^2 \gamma_2^2 f(\nu) [\cos (\pi\nu)-\cos(3\pi\nu)]V^{2\nu-1} T^{2\nu-2},  \label{resultS} \\
\delta S  &\simeq - 2{e^*}^2 \gamma_1^2 \gamma_2^2 f(\nu) [\cos(\pi\nu)-\cos(3\pi\nu)]V^{2\nu-1} T^{2\nu-2}. \nonumber
\end{align} 
This gives Eq.~\eqref{excessN}~\cite{Correction1,Supple}. Notice that $I>0$ but $\delta S < 0$. 
The factors having $\pi \nu$ originate from anyon braiding.
%$-[\cos \pi \nu - \cos 3 \pi \nu]$ in $\delta S$ originates from anyon braiding and exchange.
%$\delta S < 0$ means that the nonequilibrium noise $S(V,T)$ is smaller than the thermal noise $S(0,T)$.

The current $I$ and excess noise $\delta S$ are linked to measurable quantities.
$I$ equals the average current $I_3 = \overline{I_3(t)}$ at D3, as only S1 is biased.
$\delta S$ is obtained~\cite{Supple} by 
\begin{align}  \delta S = & \,\,\, S_3 (V, T) - 4 k_B T \dfrac{\partial  I_3 (V,T,V_3)}{\partial V_3}\bigg\vert_{V_3=0} \nonumber \\ - & \left[ S_3 (0, T) - 4 k_B T \dfrac{\partial  I_3 (0,T,V_3)}{\partial V_3}\bigg\vert_{V_3=0} \right]. \label{S3} \end{align}
The noise $S_3 (V,T) = 2 \int^\infty_{-\infty} dt (I_3(t) - I_3) (I_3(0) - I_3)$ is measured at D3. $\partial I_3 (V,T,V_3) / \partial V_3 |_{V_3 = 0}$ is measured with the voltage $V_3$ applied to S3 in addition to the voltage $V$ at S1, 
and equals the correlation between the tunneling current $I(t)$ at QPC2 and the current from S3 to QPC2, according to the nonequilibrium fluctuation-dissipation theorem~\cite{Wang1,Wang2,Smits}.
%Note that the second term of Eq.~\eqref{S3} is negligible in the case of electrons at $\nu=1$ while non-negligible for the anyon case.

{\it Main processes.---} We discuss the origin of $\delta S < 0$. The tunneling current and its excess noise satisfy~\cite{Feldman2} $I = e^* (W_{2 \to 3}-W_{3 \to 2})$ and $\delta S = 2(e^*)^2 (W_{2 \to 3} + W_{3 \to 2})$. $W_{2 \to 3}$ ($W_{3 \to 2}$) is the change,  by the voltage $V$, in the rate for a particle-like (hole-like) anyon to move from Edge2 to Edge3 at QPC2.
Two types of processes, Poisson processes and TVBs, make contribution $W^\textrm{P}_{i \to j}$ and $W^\textrm{TVB}_{i \to j}$, respectively, to $W_{i \to j}$,
\begin{equation}
W_{i \to j} \simeq W^\textrm{P}_{i \to j} + W^\textrm{TVB}_{i \to j} \quad \quad \textrm{at} \,\,\, e^*V \gg k_B T. \label{rate_total}
\end{equation}
%\begin{align} &W_{2\rightarrow 3}=W_{2\rightarrow 3 }^\textrm{P}+W_{2\rightarrow 3 }^\textrm{TVB}\nonumber \\&W_{3\rightarrow 2}=W_{3\rightarrow 2}^\textrm{P}+W_{3\rightarrow 2}^\textrm{TVB} \end{align} 
$W_{i \to j}$ is computed in Ref.~\cite{Supple}.

In the Poisson process [Fig.~\ref{setup}(b)] for $W^\textrm{P}_{2 \to 3}$, 
a particle-like anyon, biased by the voltage $V$, moves from Edge1 to Edge3 through tunneling at QPC1 and QPC2.
This leads to $W_{2 \to 3}^\textrm{P} \propto \gamma_1^2\gamma_2^2 V^{4\nu-3}$,
%\begin{equation}
%W_{2 \to 3}^\textrm{P} \propto \gamma_1^2\gamma_2^2 V^{4\nu-3}, \label{W_P}
%\end{equation}
as the voltage-biased tunneling probability at QPC$i$ and the current from S1 to QPC1 are proportional to
$\gamma_i^2V^{2\nu-2}$ and $V$, respectively.
By contrast, $W^\textrm{P}_{3 \to 2} = 0$, since tunneling of a hole-like anyon from Edge2 to Edge3 is not induced by $V$.

Next, we consider the TVB for $W^\textrm{TVB}_{3 \to 2}$.
It is the interference of two subprocesses $a_1$ and $a_2$ [Fig.~\ref{setup}(c)].
In $a_1$ and $a_2$, a particle-like anyon, induced by the voltage $V$, moves from Edge1 to Edge2 via tunneling at QPC1 at time $t_1$, and then moves to D2. The operator for the QPC1 tunneling is $\mathcal{T}_{1\rightarrow 2}(t_1) = \Psi_2^{\dagger}(0,t_1)\Psi_1(0,t_1)$. $\Psi^\dagger_i (x_i, t_1)$ creates an anyon at position $x_i$ of  Edge$i$; QPC1 is located at $x_i = 0$.
After (before) this anyon passes QPC2, a particle-hole pair is thermally excited at QPC2 at time $t_2$ ($t_2'$) in the subprocess $a_1$ ($a_2$). 
Then the particle-like thermal anyon moves to D2 along Edge2, while the hole-like one to D3 along Edge3.
The excitation is described by the QPC2 tunneling operator $\mathcal{T}_{3 \to 2}(t) = \Psi_2^\dagger (d,t) \Psi_3 (0,t)$ at $t = t_2$ ($t_2'$) in $a_1$ ($a_2$); QPC2 is located at $x_2 = d$ ($x_3 = 0$) on Edge2 (Edge3). 

To illustrate the nontrivial features (topological link by anyon braiding and the partner disconnected process) of the TVB for $W^\textrm{TVB}_{3 \to 2}$,
we consider the $V \to \infty$ limit where the voltage-biased particle-like anyon becomes a point particle (its spatial broadening $\hbar v / (e^*V) \to 0$; $v$ is the anyon velocity). In this limit,  the correlator
\begin{align}
C_{3\rightarrow 2}^\textrm{TVB} =   &\langle  \mathcal{T}_{1\rightarrow 2}^\dagger (t_1) \mathcal{T}_{3\rightarrow 2}^\dagger (t_2') \mathcal{T}_{3\rightarrow 2}(t_2) \mathcal{T}_{1\rightarrow 2}(t_1)  \rangle \nonumber \\ 
 -& \langle \mathcal{T}_{3\rightarrow 2}^\dagger (t_2') \mathcal{T}_{3\rightarrow 2} (t_2) \rangle \langle \mathcal{T}_{1\rightarrow 2}^\dagger (t_1)  \mathcal{T}_{1\rightarrow 2} (t_1)\rangle \label{correlator}
\end{align}
describes the TVB. $\langle \cdots \rangle$ is the ensemble average with the bare Hamiltonian~\cite{Supple} $H_i$ of Edge$i$. %This correlator appears in the expression~\cite{Supple} of $W_{i \to j}$.

\begin{figure}[bht] 
\centering
\includegraphics[width=\linewidth]{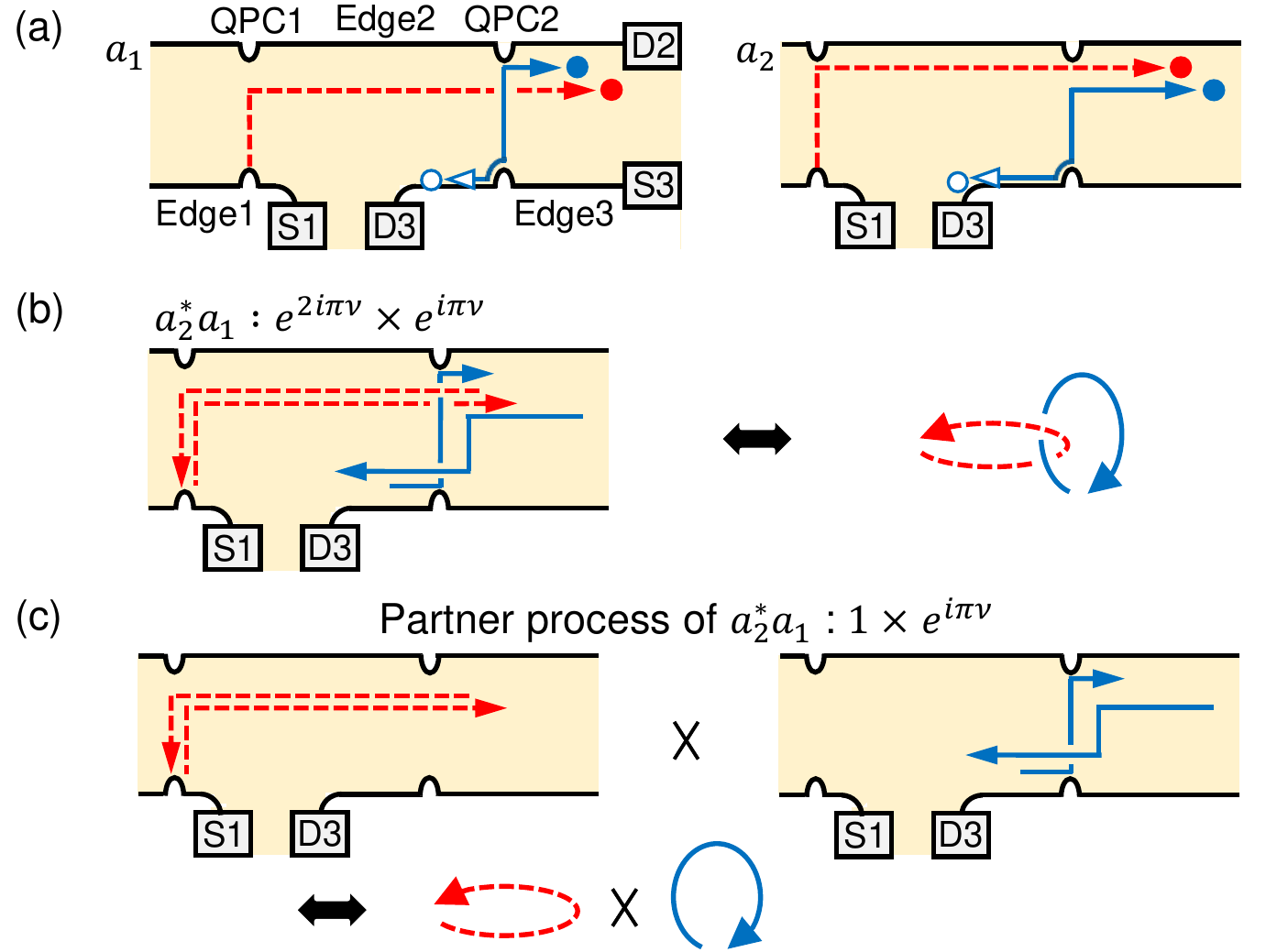}  
\caption{TVB interference for $W^\textrm{TVB}_{3 \to 2}$. (a) Its subprocesses $a_1$ and $a_2$ [identical to those in Fig.~\ref{setup}(c)] have the trajectory (dashed red arrows) of a voltage-biased anyon (red filled circles)
and that (solid blue) of a thermal pair excitation of a particle-like anyon (blue filled) and a hole-like anyon (blue empty).
Two trajectories are drawn to cross when the corresponding operators are non-commutative due to the fractional statistics.
The crossing is time-ordered such that the later trajectory is drawn on top of the earlier one. 
(b) TVB interference $a_2^* a_1$ between $a_1$ and $a_2$.
The trajectories of $a_2^*$, the complex conjugation of $a_2$, are drawn on top of those of $a_1$.
%; the trajectory directions of the complex-conjugation $a_2^*$ are opposite to those of $a_2$.
The loop formed by the dashed red trajectories is topologically linked with that by the solid blue ones, implying effective braiding of the thermal anyon around the voltage-biased anyon. The braiding phase factor is $e^{ 2i \pi \nu}$.
(c) In the partner disconnected process of $a_2^* a_1$, the two loops are unlinked, showing no braiding.
$a_2^* a_1$ and its partner have a common phase factor $e^{i \pi \nu}$ due to an exchange of a thermal anyon of $a_1$ and another of $a_2$ (the crossing of solid blue trajectories).
} \label{process} 
\end{figure} 

The first term of Eq.~\eqref{correlator} shows the interference between the subprocesses $a_1$ and $a_2$; 
$\mathcal{T}_{3 \to 2}(t_2) \mathcal{T}_{1 \to 2}(t_1)$ describes $a_1$, while $\mathcal{T}_{3 \to 2} (t_2') \mathcal{T}_{1 \to 2} (t_1)$ describes $a_2$.
This term is factorized~\cite{Supple} into a subcorrelator for the voltage-biased anyon, another for the thermal anyons, and a phase factor $e^{i 2 \pi \nu}$ (Fig.~\ref{process}),
\begin{align}
& \langle \mathcal{T}_{1\to 2}^\dagger (t_1) \mathcal{T}_{3\to 2}^\dagger (t_2') \mathcal{T}_{3\to 2}(t_2) \mathcal{T}_{1\to 2}(t_1)  \rangle \nonumber \\
& =  e^{2i\pi\nu} \langle \mathcal{T}_{3\to 2}^\dagger (t_2') \mathcal{T}_{3\to 2}(t_2) \rangle \langle \mathcal{T}_{1\to 2}^\dagger (t_1)  \mathcal{T}_{1\to 2}(t_1) \rangle, \label{braiding}
\end{align}
by using the exchange rules of the fractional statistics $\Psi_i^{\dagger}(x) \Psi_i (y) = \Psi_i (y) \Psi_i^{\dagger} (x) e^{i\pi\nu \textrm{sgn} (x-y)}$ and $\Psi_i^{\dagger} ( x) \Psi_i^{\dagger}(y) = \Psi_i^{\dagger} (y) \Psi_i^{\dagger} (x) e^{-i\pi\nu \textrm{sgn} (x-y )}$ (the rules between operators of different edges are constructed,
%in the connected edge scheme, 
using Klein factors~\cite{Guyon,Supple}). 
The factor $e^{i 2 \pi \nu}$ is attributed to effective braiding %~\cite{Supple,DoubleE} 
of the thermal anyon around the voltage-biased anyon in the interference $a_2^* a_1$, depicted as the link of two loops in Fig.~\ref{process}(b); the factorization is equivalent to untying the link.
The solid blue loop corresponding to the subcorrelator $\langle \mathcal{T}_{3\to 2}^\dagger (t_2') \mathcal{T}_{3\to 2}(t_2) \rangle$ for the thermal anyons is formed, although $t_2 \ne t_2'$, with the help of the thermal length $\hbar v / (k_B T)$; $\langle \mathcal{T}_{3\to 2}^\dagger (t_2') \mathcal{T}_{3\to 2}(t_2) \rangle$ is nonvanishing for $|t_2 - t_2'| \lesssim \hbar / (k_B T)$. Similarly, at finite $V$, the dashed red loop representing $\langle \mathcal{T}_{1\to 2}^\dagger (t_1)  \mathcal{T}_{1\to 2}(t_1) \rangle$ for the voltage-biased anyon is formed with $|t_1 - t_1'| \lesssim \hbar v / (e^* V)$, when the tunneling at QPC1 occurs at $t_1'$ ($\ne t_1$) in $a_2$ as described by $\mathcal{T}_{1\rightarrow 2} (t_1')$.
In this case, the braiding occurs for $t_2' <  t_1 + d/v < t_2$ and  $t_2' <  t_1' + d/v < t_2$.

 The effective braiding ($e^{2i\pi\nu}$) is decomposed into two events of anyon exchange.  One exchange ($e^{i\pi\nu}$) occurs in the subprocess $a_1$ when the thermal anyon is excited on Edge2 at QPC2 [Fig.~\ref{process}(a)]. It happens such that the thermal anyon effectively moves from the right side of the voltage-biased anyon to the left on Edge2~\cite{Supple}. The other ($e^{i\pi\nu}$) occurs in the interference $a_2^* a_1$. The voltage-biased anyon of $a_2$ moves back to QPC1 passing the thermal anyon of $a_1$ [the top dashed arrow in Fig.~\ref{process}(b)].

We call the first term of Eq.~\eqref{correlator} a TVB since 
%a Feynmann diagram whose subdiagrams, the dashed red loop of the voltage-biased anyon and the solid blue loop of the thermal anyon, are disconnected to each other in the conventional sense but topologically linked by the braiding.
the trajectory (dashed red loop) of the voltage-biased anyon and that (solid blue loop) of the thermal anyon are disconnected to each other in the conventional sense but topologically linked~\cite{AlgebraicT} by the braiding. The TVB is accompanied by a partner {\em disconnected} process [Fig.~\ref{process}(c)] that gives the second term of Eq.~\eqref{correlator} and has the same subprocesses as the TVB except the braiding.
The TVB and its partner disconnected process (or the correlator in Eq.~\eqref{correlator}) appear in our calculation~\cite{Supple} of $W_{i \to j}$. 
The pairwise appearance is understood by considering electrons at $\nu=1$.
For the electrons, the TVB is described by a disconnected Feynman diagram as the braiding link has no meaning, $e^{2 i \pi \nu} = 1$. Then it must be accompanied and exactly cancelled (leading to $C_{3 \to 2}^\textrm{TVB} = 0$; cf. Eqs.~\eqref{correlator} and \eqref{braiding}) by the partner disconnected diagram, following the linked cluster theorem~\cite{Fetter}; the second term of Eq.~\eqref{correlator} has the minus sign for the cancellation; mathematically, the partner diagram appears due in part to the partition function of a Green's function in its perturbation expansion, hence it does not have the braiding link.
For the anyons, the cancellation is partial, because of the braiding. 
%The TVB and its partner disconnected processes together determine $W_{3 \to 2}^\textrm{TVB}$.

The common factor of the two terms of Eq.~\eqref{correlator} is further factorized with
a correlator  $D_i (x, t, t') = \langle \Psi^\dagger_i (x,t) \Psi_i (x,t')\rangle$ of each Edge$i$,
\begin{align}
& \langle \mathcal{T}_{3\to 2}^\dagger(t_2') \mathcal{T}_{3\to 2} (t_2) \rangle \langle \mathcal{T}_{1\to 2}^\dagger (t_1)  \mathcal{T}_{1\to 2} (t_1) \rangle \label{exchange_thermal} \\
& = e^{i\pi\nu} D_2(d,t_2,t_2') D_3(0,t_2,t_2') D_1 (0,t_1,t_1) D_2(0, t_1,t_1). \nonumber
\end{align}
%$\langle \mathcal{T}_{2\to 3}^\dagger(t_2') \mathcal{T}_{2\to 3} (t_2) \rangle \langle \mathcal{T}_{1\to 2}^\dagger (t_1)  \mathcal{T}_{1\to 2} (t_1) \rangle = e^{i\pi\nu}\langle \Psi^{\dagger}_2 (d,t_2 ) \Psi_2 (d,t_2' ) \rangle \langle \Psi^{\dagger}_3 (0,t_2) \Psi_3 (0,t_2' ) \rangle \langle \Psi^{\dagger}_1 (0,t_1 ) \Psi_1 (0,t_1) \rangle \langle \Psi^{\dagger}_2 (0,t_1) \Psi_2 (0,t_1) \rangle$
%\begin{widetext} \begin{align}
% & \langle \mathcal{T}_{2\to 3}^\dagger(t_2') \mathcal{T}_{2\to 3} (t_2) \rangle \langle \mathcal{T}_{1\to 2}^\dagger (t_1)  \mathcal{T}_{1\to 2} (t_1) \rangle \nonumber \\
% & = \langle\Psi^{\dagger}_2 (d, t_2') \Psi_3 (0,t_2') \Psi^{\dagger}_3 (0,t_2)\Psi_2 (d,t_2 ) \rangle \langle \Psi^{\dagger}_1 (0,t_1) \Psi_2 (0,t_1) \Psi^{\dagger}_2 (0,t_1) \Psi_1 (0,t_1) \rangle \nonumber \\
% & = e^{i\pi\nu}\langle \Psi^{\dagger}_2 (d,t_2 ) \Psi_2 (d,t_2' ) \rangle \langle \Psi^{\dagger}_3 (0,t_2) \Psi_3 (0,t_2' ) \rangle \langle \Psi^{\dagger}_1 (0,t_1 ) \Psi_1 (0,t_1) \rangle \langle \Psi^{\dagger}_2 (0,t_1) \Psi_2 (0,t_1) \rangle 
%\end{align} \end{widetext} 
The factor $e^{i \pi \nu}$ comes from exchange of a thermal anyon of $a_1$ and another of $a_2$ [Figs.~\ref{process}(b,c)].

The TVB and its partner disconnected process give 
\begin{equation}
W_{3 \to 2}^\textrm{TVB} \propto  \gamma_1^2\gamma_2^2 V^{2\nu-1}T^{2\nu-2} \textrm{Re} [e^{i \pi \nu} (e^{2i \pi \nu} - 1)], \label{W_TVB}
\end{equation}
as the thermal (voltage-biased) tunneling probability at QPC2 (QPC1) is proportional to $\gamma_2^2 T^{2\nu-2}$ ($\gamma_1^2V^{2\nu-2}$) while the current from S1 to QPC1 is proportional to $V$.
The phase factors come from $\textrm{Re} [C_{3\rightarrow 2}^\textrm{TVB}] \propto \textrm{Re}[ e^{i \pi \nu} (e^{2i \pi \nu} - 1)]$ in Eqs.~\eqref{correlator}-\eqref{exchange_thermal}. $\textrm{Re} [\cdots]$ is taken, considering
% since there is also 
$[C_{3\rightarrow 2}^\textrm{TVB}]^*$.  
   
There is a TVB process for $W^\textrm{TVB}_{2 \to 3}$.
$W^\textrm{TVB}_{2 \to 3}$ is negligibly small at $e^*V \gg k_B T$~\cite{W23}.

%In addition, there is a TVB process that determines $W^\textrm{TVB}_{2 \to 3}$. This process is 
%%the interference of $a_1 = \mathcal{T}_{2 \to 3}^\dagger (t_2) \mathcal{T}_{1 \to 2}(t_1)$ and $a_2 =  \mathcal{T}_{2 \to 3}^\dagger (t_2') \mathcal{T}_{1 \to 2} (t_1')$ [Figs.~\ref{process}(c) and \ref{setup}(c)].  
%identical to that of $W^\textrm{TVB}_{3 \to 2}$, except that  its thermal anyons move in the opposite direction to those of  $W^\textrm{TVB}_{3 \to 2}$; the particle-like (hole-like) thermal anyon is excited on Edge3 (Edge2) and moves to D3 (D2).  Because of the difference,  the TVB for $W^\textrm{TVB}_{2 \to 3}$ has the braiding phase factor $e^{-2 i \pi \nu}$, and $W^\textrm{TVB}_{2 \to 3}$ has the same expression as $W^\textrm{TVB}_{3 \to 2}$ in Eq.~\eqref{W_TVB}, but with the replacement of $e^{2 i \pi \nu} \to e^{-2 i \pi \nu}$. $W^\textrm{TVB}_{2 \to 3}$ is negligibly small at $e^*V \gg k_B T$ as $W^\textrm{TVB}_{2 \to 3} \propto \textrm{Re}[e^{i \pi \nu} (e^{-2i \pi \nu} - 1)] = 0$.
    
We now compute $\delta S/I$.
At $e^* V \gg k_B T$ and $\nu = 1/(2n+1) < 1$, the TVB for $W_{3 \to 2}^\textrm{TVB}$ and its partner disconnected process dominate over the Poisson process for $W_{2 \to 3}^\textrm{P}$,  $W_{3 \to 2}^\textrm{TVB} \gg W_{2 \to 3}^\textrm{P}$; cf. Eq.~\eqref{W_TVB} and $W_{2 \to 3}^\textrm{P} \propto \gamma_1^2\gamma_2^2 V^{4\nu-3}$. Hence, they determine the current and the excess noise,
$I = - e^* W_{3 \to 2}^\textrm{TVB}$ and $\delta S = 2 (e^*)^2 W_{3 \to 2}^\textrm{TVB}$, leading to Eqs.~\eqref{excessN} and \eqref{resultS}.
We emphasize that the ratio $\delta S / I$ has the negative universal value of $- 2 e^*$. This originates from the TVB for $W_{3 \to 2}^\textrm{TVB}$ and its partner disconnected process, and equivalently from the anyon braiding. It is nontrivial that the disconnected process contributes to the observables $I$ and $\delta S$; for electrons or bosons, disconnected Feynman diagrams never contribute to observables~\cite{Fetter}.

The above findings are confirmed by numerically computing $\delta S$~\cite{Supple}.
For $\nu = 1/3$, $\delta S$ approaches to $- 2 e^* I$ such that $\delta S = - 1.8 e^* I$ at $V = 60$ $\mu$V at 50 mK
and $- 1.99 e^* I$ at 80 $\mu$V at 50 mK.
 
{\it Discussion.---} The negative excess noise $\delta S < 0$ results from the TVB process for $W_{3 \to 2}^\textrm{TVB}$. It is interpreted as follows. At $V=0$, tunneling of a particle-like or hole-like anyon between Edge2 and Edge3 is thermally induced at QPC2, causing the thermal noise $S(0,T)$. 
Among those tunneling events, thermal tunneling of a hole-like anyon from Edge2 to Edge3 is weakened by a voltage-biased particle-like anyon injected from Edge1 to Edge2, when  the voltage $V$ is applied to Edge1.
The weakening is due to the effective braiding of the thermal anyon around the voltage-biased anyon, which results in
the partial cancellation between the TVB and its partner disconnected process, $W_{3 \to 2}^\textrm{TVB} \propto \textrm{Re} [e^{i \pi \nu} (e^{2i \pi \nu} - 1)] < 0$.
The weakening leads to the current $I > 0$ and the reduction of the noise $S(V,T)$ below $S(0,T)$. 
Note that $\delta S < 0$ at any $V$, although both the Poisson process and the TVB (and its partner) contribute to $\delta S$ at $e^*V \lesssim k_B T$.

By contrast, for electrons at $\nu=1$, the Poisson process determines $I = e W_{2 \to 3}^\textrm{P}$ and $\delta S = 2 e^2 W_{2 \to 3}^\textrm{P}$, leading to $\delta S = 2 e I>0$ at $e^* V \gg k_B T$. There is no topological link by the braiding ($e^{2i \pi \nu} = 1$), and the TVB becomes a disconnected process and fully cancelled by its partner disconnected diagram, $W_{i \to j}^\textrm{TVB}= 0$.  
This is why the excess noise $\delta S = 2 e I$ of the electrons cannot be extrapolated from Eq.~\eqref{excessN} with $e^* \to e$.
 
%  topological vacuum bubbles, a new type of anyon transport described by processes disconnected in the conventional sense but topologically linked~\cite{AlgebraicT} by anyon braiding.
  
%The excess noise is negative for any bias voltage as you see in Fig 1. (b) but universal value $S=-2e^*I$ can be obtained in the large bias voltage regime. (at least $e^*V>4.5k_BT$ will guarantee $<$10\% error. when $T=50mK$, $V>60\mu V$ will guarantee)

Measurement of $\delta S$ is feasible, as the setup was experimentally studied in other contexts~\cite{Comforti,Chung,Comforti2}:
Typically, the tunneling probability of QPC1 and QPC2 is set to be 0.2, to have anyon tunneling~\cite{Picciotto}. We estimate $I \sim 50$ pA and $\delta S \sim$  $2.7 \times 10^{-30}$ A$^2$/Hz at 100 $\mu$V and $\nu = 1/3$, which is detectable~\cite{Comforti,ChungChoi}. 
When $\delta S$ is measured by using Eq.~\eqref{S3}, one has to experimentally determine temperature $T$. 
The determination accuracy is within $\pm 3$ mK~\cite{ChungChoi}. Then, it is possible to obtain $\delta S = - 2 e^* I (1 \pm 0.2)$ at 50 mK, $V = 80$ $\mu$V, and $\nu = 1/3$.

Our study is generalized to edges with multiple channels or reconstruction (see Ref.~\cite{Supple}).
 For example, at filling factor $4/3$ or $7/3$~\cite{An,Baer},  the inner fractional edge channel corresponding to $\nu =1 /3$ interacts with co-propagating outer channels, and is weakly backscattered at the QPCs. In this case $\delta S$ is still negative.
 % due to the fractional statistics, 
On the other hand, when the $\nu=1/3$ edge channel interacts with an unexpected counter-propagating mode~\cite{Rosenow_edge_reconst} due to edge reconstruction, $\delta S$ is negative only when the interaction is sufficiently weak~\cite{Saminadayar,Roddaro}.
The outer channels at filling factor $4/3$ or $7/3$ are helpful in this case, since they can screen the edge reconstruction.
In the above cases of multiple channels or edge reconstruction, detection of $\delta S < 0$ may imply the fractional statistics of the quasiparticles deviating from Laughlin anyons due to the interchannel interactions. The quasiparticles become closer to Laughlin anyons for weaker interactions.
 
% The outer channels are helpful for detecting the negative excess noise, since they can screen unexpected edge reconstruction;  interaction of the inner channel with an upstream mode induced by edge reconstruction~\cite{Rosenow_edge_reconst} is harmful for detecting the negative excess noise, but the excess noise is negative for weak interaction~\cite{Saminadayar,Roddaro} (see Ref.~\cite{Supple} for the details).  
 
%And, the inter-QPC distance $d$ on Edge2 should be less than energy relaxation lengths, to prevent the voltage-biased anyon from energy loss. The intraedge relaxation length exceeds hundreds of microns at $\nu=1/3$~\cite{Comforti2}.

In summary, we predict the negative excess autocorrelation noise $\delta S <0$, a signature of the Abelian fractional statistics or the new process (TVB) not existing with fermions or bosons. 
It is unusual that the excess autocorrelation noise of electrical tunneling current is negative~\cite{Blanter,Lesovik}.

We suggest that autocorrelation noise can provide signatures~\cite{Buttiker,Lee} of identical-particle statistics. This is different from the conventional approach~\cite{Henny,Oliver,Jeltes} of detecting particle bunching or antibunching with Hanbury Brown-Twiss cross-correlations.
It is unnatural to interpret the negative excess autocorrelation noise as deviation (anyonic partial bunching~\cite{Safi,Vishveshwara,Kim06,Campagnano12,Rosenow2}) from fermionic antibunching and bonsonic bunching, because it originates from the TVB having no counterpart in fermions or bosons.
 
%{\it Appendix.---}  The Hamiltonian of the setup is $H = \sum_i H_{i} + H_\textrm{tun}$. Edge$i$ is described by the chiral Luttinger liquid~\cite{Wen,vonDelft} $H_{i} =\frac{\hbar v}{4\pi \nu} \int_{-\infty}^{\infty} dx :(\partial_x\phi_i(x))^2:+e^* V_i \hat{N}_i$, the anyon operator $\Psi^\dagger (x,t) = F_i^\dagger (t) e^{-i \phi_i (x,t)} / \sqrt{2 \pi a}$ by the boson field $\phi_i (x)$, and anyon tunneling at the QPCs by $H_\textrm{tun} = \gamma_1 (\mathcal{T}_{1 \to 2}+ \mathcal{T}^\dagger_{1 \to 2}) +  \gamma_2 ( \mathcal{T}_{2 \to 3} + \mathcal{T}^\dagger_{2 \to 3})$. Here, $V_1 = V$, $V_{2,3} = 0$, $\hat{N}_i$ is the anyon number operator obeying $[N_i. F_j^\dagger] = \delta_{ij} F_i^\dagger$, $a$ is the short-length cutoff, and $F^\dagger_i$ is the Klein factor~\cite{Guyon} satisfying $F^\dagger_i F_i = 1$, $F_1(t) = F_1(0) e^{-i e^* V t / \hbar}$, and $[\phi_i, F_j]=0$. We treat $H_\textrm{tun}$ as a perturbation on $\sum_i H_i$. The operators for $I$ and $S$ at QPC2 are  $\hat{I} = i e^* \gamma_2 ( \mathcal{T}_{2 \to 3} - \mathcal{T}^\dagger_{2 \to 3})/\hbar$, $\hat{S} =  \int_{-\infty}^{\infty} dt \delta \hat{I}(0) \delta \hat{I} (t) + \delta \hat{I}(t) \delta \hat{I} (0)$, and $\delta \hat{I} (t) = \hat{I} (t) -  I$.  

%We compute $W_{i \to j}$, using the Keldysh method. The expression of $W_{i \to j}$ and its detailed computation are found in Ref.~\cite{Supple}. The correlator in Eq.~\eqref{correlator} appears in the expression of $W_{i \to j}$.

We thank Hyungkook Choi, Sang-Jun Choi, Yunchul Chung, Sourin Das, Dmitri Feldman, Bertrand Halperin, Charles Kane, and Bernd Rosenow for valuable discussions, and the support by Korea NRF (SRC Center for Quantum Coherence in Condensed Matter, Grant No. 2016R1A5A1008184).


\begin{thebibliography}{}
\bibitem{Correction1} Note that the correction of order $O (T/V)$ to Eq.~\eqref{excessN} is found as
$\delta S / I = - 2 e^* [ 1 + \frac{4 \nu (1-2\nu)}{\sin 2 \pi \nu} \frac{\pi k_B T}{e^* V}]$. It occurs due to the spatial broadening $\hbar v / (e^* V)$ of the voltage-biased anyon that leads to imperfection of the effective braiding.
\bibitem{W23} The TVB process for $W^\textrm{TVB}_{2 \to 3}$ is identical to that of $W^\textrm{TVB}_{3 \to 2}$, except that its thermal anyons move in the opposite direction to those of  $W^\textrm{TVB}_{3 \to 2}$.
%; the particle-like (hole-like) thermal anyon is excited on Edge3 (Edge2) and moves to D3 (D2).  Because of the difference,  the TVB for $W^\textrm{TVB}_{2 \to 3}$ has the braiding phase factor $e^{-2 i \pi \nu}$, and 
$W^\textrm{TVB}_{2 \to 3}$ has the same expression as $W^\textrm{TVB}_{3 \to 2}$ in Eq.~\eqref{W_TVB}, but with the replacement of $e^{2 i \pi \nu} \to e^{-2 i \pi \nu}$.
\end{thebibliography}
\end{document}

\renewcommand{\theequation}{S\arabic{equation}}
\renewcommand{\thefigure}{S\arabic{figure}}

\title{Supplemental Material : Negative excess shot noise by anyon braiding}
\date{\today}
\author{Byeongmok Lee}
\affiliation{Department of Physics, Korea Advanced Institute of Science and Technology, Daejeon 34141, Korea}
\author{Cheolhee Han}
\affiliation{Department of Physics, Korea Advanced Institute of Science and Technology, Daejeon 34141, Korea}
\author{H.-S. Sim}
\email[Corresponding author. ]{hssim@kaist.ac.kr}
\affiliation{Department of Physics, Korea Advanced Institute of Science and Technology, Daejeon 34141, Korea}
\maketitle

This material includes the Hamiltonian and operators of the setup, the Klein factors, the expression of $f(\nu)$, the derivation of Eqs.~(3) and (6), the effective braiding, the detailed expression and computation of $W_{i \to j}$, the correction of order $O(T/V)$ to Eq.~(1), and the effect of multiple edge channels and edge reconstruction.

\subsection{Hamiltonian and operators of the setup, and Klein factors}

The Hamiltonian of the setup is 
\begin{eqnarray} H = \sum_i H_{i} + H_\textrm{tun}. \label{totH_o} \end{eqnarray} 
Edge$i$ is described by the chiral Luttinger liquid~\cite{Wen,vonDelft} $H_{i} =\frac{\hbar v}{4\pi \nu} \int_{-\infty}^{\infty} dx :(\partial_x\phi_i(x))^2:+e^* V_i \hat{N}_i$, the anyon operator $\Psi^\dagger_i (x,t) = F_i^\dagger (t) e^{-i \phi_i (x,t)} / \sqrt{2 \pi a}$ by the boson field $\phi_i (x)$, and anyon tunneling at the QPCs by $H_\textrm{tun} = \gamma_1 (\mathcal{T}_{1 \to 2}+ \mathcal{T}^\dagger_{1 \to 2}) +  \gamma_2 ( \mathcal{T}_{2 \to 3} + \mathcal{T}^\dagger_{2 \to 3})$. Here, $V_1 = V$, $V_{2,3} = 0$, $\hat{N}_i$ is the anyon number operator obeying $[N_i. F_j^\dagger] = \delta_{ij} F_i^\dagger$, $a$ is the short-length cutoff, and $F^\dagger_i$ is the Klein factor~\cite{Guyon} satisfying $F^\dagger_i F_i = 1$, $F_1(t) = F_1(0) e^{-i e^* V t / \hbar}$, and $[\phi_i, F_j]=0$. The exchange rule between anyons on different edges is achieved with the commutators of $F_i$, 
\begin{align}
&F_1 F_2 =F_2 F_1 e^{i \pi\nu }, \quad \quad F_2 F_3 =F_3 F_2 e^{i \pi\nu }, \quad \quad F_1 F_3 =F_3 F_1 e^{i \pi\nu },   \nonumber \\
&F_1^\dagger F_2 =F_2 F_1^\dagger e^{-i \pi\nu}, \quad F_2^\dagger F_3 =F_3 F_2^\dagger e^{-i \pi\nu}, \quad F_1^\dagger F_3 =F_3 F_1^\dagger e^{-i \pi\nu}.
\label{Klein_commute} \end{align} 
We treat $H_\textrm{tun}$ as a perturbation on $\sum_i H_i$.
The operators for $I$ and $S$ at QPC2 are 
$\hat{I} = i e^* \gamma_2 ( \mathcal{T}_{2 \to 3} - \mathcal{T}^\dagger_{2 \to 3})/\hbar$, $\hat{S} =  \int_{-\infty}^{\infty} dt \delta \hat{I}(0) \delta \hat{I} (t) + \delta \hat{I}(t) \delta \hat{I} (0)$, and $\delta \hat{I} (t) = \hat{I} (t) -  I$.

\begin{figure}[b] 
\centering
\includegraphics[width=.9\columnwidth]{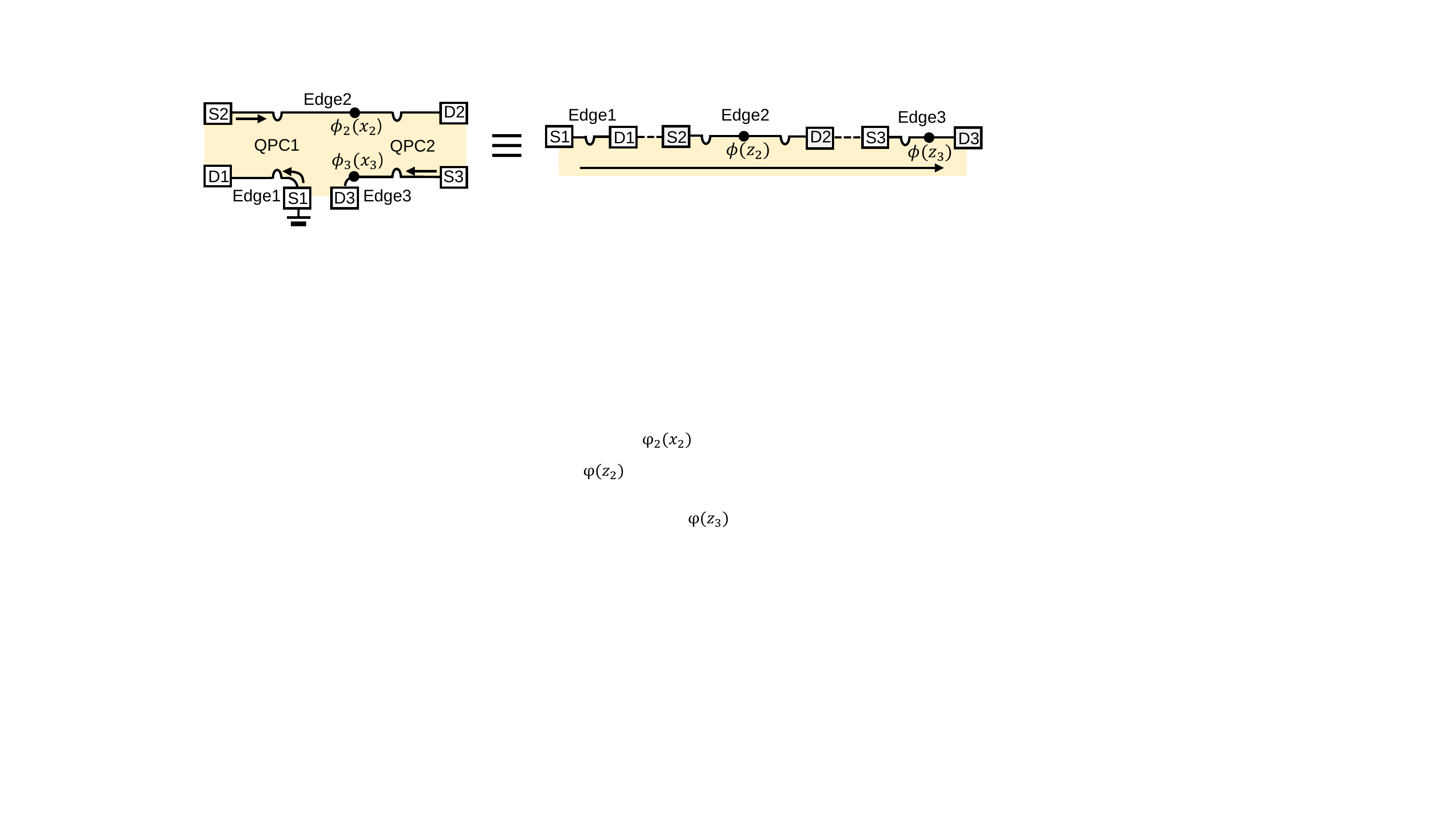}
\caption{A single connected edge (the right panel) equivalent with our setup (the left). The dashed line between D1 and S2 in the right panel represents the edge connecting Edge1 and Edge2, and the dashed line between D2 and S3 shows the edge connecting Edge2 and Edge3. The connected edge has the same chirality (see the arrows) with Edge1, Edge2, and Edge3.  }  \label{connected_edge} 
\end{figure} 

%Klein factors are essential when one calculates transport quantities in setups with multiple edges.
We attach the Klein factors in Eq.~\eqref{Klein_commute}, utilizing the connected-edge scheme~\cite{Guyon}. The connected edge is constructed by gluing Edge1, Edge2, and Edge3, and it has the same chirality with Edge$i$'s. 
Here, Edge$i$'s are regarded as part of the boundary (which is the connected edge) of one fractional quantum Hall liquid.
Then the commutation rule between the boson field $\phi_2 (x_2)$ of Edge2 and the boson field $\phi_3 (x_3)$ of Edge3, for example, can be assigned, following the commutation rule $\left[ \phi (z_2), \phi (z_3) \right] = i \pi \nu \textrm{sgn}(z_2-z_3)$ of the corresponding boson field $\phi$ of the connected edge of coordinate $z$  (see Fig.~\ref{connected_edge}). Similarly, the commutation rule between the quasiparticle operators of two different edges can be assigned. This leads to Eq.~\eqref{Klein_commute}.
 %Klein factors ensure that the quasiparticle operators in differenct edges follow correct exchange rules. The commutators of Klein factors are determined in such a way that the exchange rules between quasiparticle of different edges are same as the one with a closed system where edges are connected with an open contour which brings all the edges into one.(Fig connected edge)
 
We note that for our setup, one can omit the Klein factors in the limiting case where the distance between QPC1 and QPC2 in Edge2 is infinite (much longer than the spatial broadening of thermal anyons and voltage-biased anyons). In this case, it is enough to compute the correlator of the form
$\langle  \mathcal{T}_{1\rightarrow 2}^\dagger (t_1) \mathcal{T}_{3\rightarrow 2}^\dagger (t_2') \mathcal{T}_{3\rightarrow 2}(t_2) \mathcal{T}_{1\rightarrow 2}(t_1)  \rangle$  (see Eq. (5) in the main text) for the calculation of $I$ and $\delta S$. In the computation of the correlator,  the Klein factor occuring in  $\mathcal{T}_{3\rightarrow 2}(t_2) \mathcal{T}_{1\rightarrow 2}(t_1)$ is the same with, hence cancels exactly, that in $ \mathcal{T}_{3\rightarrow 2} (t_2') \mathcal{T}_{1\rightarrow 2} (t_1) $. Hence the same computation result is obtained without introducing the Klein factors. 

On the other hand, however, when the distance between QPC1 and QPC2 is short, the correlators of the other forms, such as $\langle \mathcal{T}_{3\rightarrow 2}^\dagger (t_2')  \mathcal{T}_{1\rightarrow 2}^\dagger (t_1)  \mathcal{T}_{3\rightarrow 2}(t_2) \mathcal{T}_{1\rightarrow 2}(t_1)  \rangle$, contribute to $I$ and $\delta S$, and the calculation of the correlators requires the Klein factors. The fractional exchange phase $e^{i \pi \nu}$ of the Klein factors in Eq.~\eqref{Klein_commute} and that of anyons on the same edges together result in the negative excess noise. To distinguish the contribution of the fractional exchange phase of the Klein factors to $I$ and $\delta S$ from that of anyon exchange on the same edge, we substitute $\nu$'s in Eq.~\eqref{Klein_commute} with $\nu_{K}$ and recalculate $I$ and $\delta S$ for the limiting case where the distance between QPC1 and QPC2 goes to zero,
\begin{align}
 I & \simeq e^* \gamma_1^2 \gamma_2^2 f(\nu) [\cos (\pi\nu - \pi (\nu + \nu_{K}))-\cos(\pi\nu + \pi (\nu + \nu_{K}) )]V^{2\nu-1} T^{2\nu-2},  \label{result_klein} \\
\delta S  &\simeq - {e^*}^2 \gamma_1^2 \gamma_2^2 f(\nu) [2\cos(\pi\nu)-\cos (\pi\nu - \pi (\nu + \nu_{K}))-\cos(\pi\nu + \pi (\nu + \nu_{K}) )]V^{2\nu-1} T^{2\nu-2}. \nonumber
\end{align}
The above equation is identical to Eq. (2) of the main text if one replaces $\nu_K$ by $\nu$. One has a wrong result if one omits the exchange phase of the Klein factors  in the case that the distance between the two QPCs is finite.
The equation~\eqref{result_klein} shows that the exchange phase $\pi \nu_K$ of the Klein factors constitutes one half of the braiding phase of the topological vacuum bubbles (TVBs), while the exchange phase of anyons on the same edge results in the other half.
Therefore, one needs to introduce the Klein factors in the general situations of our setup.

\subsection{$f(\nu)$ in Eq.~(2)}

In Eq. (2) in the main text, $f(\nu)=  \pi^{2\nu - 9/2} {e^*}^{2\nu - 1} v^{-4\nu} \hbar^{-4 \nu -1} {(k_B a^2)}^{2\nu-2} \cot(\pi\nu) \Gamma(\nu) \Gamma(1/2-\nu)  / [16 \Gamma(2\nu)]$. $\Gamma(\nu)$ is the Gamma function.
 
\subsection{Derivation of Eq.~(3)}
 
We derive Eq. (3) in the main text. The current at D3 satisfies $I_3 = I_\textrm{in} + I$ due to current conservation. $I_\textrm{in} = \overline{I_\textrm{in}(t)}$ is the average current from S3 to QPC2 along Edge3. The noise at D3 is decomposed as $S_3 = S + 2 S_{3,\textrm{in}} + S_{\textrm{in}}$, where $S_{\textrm{in}} = 2 \int^\infty_{-\infty} dt \delta I_\textrm{in} (t) \delta I_\textrm{in}(0)$, $S_{3,\textrm{in}} = 2 \int^\infty_{-\infty} dt \delta I_\textrm{in} (t) \delta I(0)$, and $\delta I_\textrm{in}(t) = I_\textrm{in}(t) - I_\textrm{in}$.
$S_\textrm{in}$ is the half of the Johnson-Nyquist noise  $4k_BT \nu e^2 / h$ and independent of $V$, hence, it does not appear in Eq. (3) in the main text. $2 S_{3,\textrm{in}}$ is  the correlation between the tunneling current $I$ at QPC2 and the current $ I_\textrm{in}$ from S3 to QPC2 and written as the second or fourth term of Eq. (3) in the main text, according to the nonequilibrium fluctuation-dissipation theorem~\cite{Wang1,Wang2,Smits}.

\subsection{Derivation of Eq.~(6)}

In the derivation of Eq. (6) in the main text, we used the identity~\cite{vonDelft} $\langle e^{s_1 \phi_1} e^{s_2 \phi_2} \cdots \rangle = e^{\sum_j s^2_j \langle \phi_j^2 \rangle + \sum_{i<j} s_i s_j \langle \phi_i \phi_j \rangle}$ with $s_i = \pm i$.

\subsection{Expression and computation of $W_{i \to j}$}  

Using the Keldysh method~\cite{Safi, Kim06}, $W_{i \to j}$ is expressed as   
\begin{align}
W_{3 \to 2} & = - \gamma_1^2 \gamma_2^2 \sum_{\eta_3 \eta_4=\pm1}\eta_3 \eta_4 \textrm{Re}[ \int_{-\infty}^\infty dt_2' dt_1 dt_1'  \nonumber \\ & \langle T_\textrm{K} \mathcal{T}_{3\to 2}(t_2^{\eta_1=1}) \mathcal{T}^\dagger_{3\to 2}(t_2'^{\eta_2=-1}) \mathcal{T}_{1\to 2}(t_1^{\eta_3}) \mathcal{T}^\dagger_{1\to 2}(t_1'^{\eta_4}) \rangle] \nonumber \\
W_{2 \to 3} & = - \gamma_1^2 \gamma_2^2 \sum_{\eta_3 \eta_4=\pm1}\eta_3 \eta_4 \textrm{Re}[ \int_{-\infty}^\infty dt_2' dt_1 dt_1'  \nonumber \\ & \langle T_\textrm{K} \mathcal{T}^\dagger_{3\to 2}( t_2^{\eta_1=1}) \mathcal{T}_{3\to 2}(t_2'^{\eta_2=-1}) \mathcal{T}^\dagger_{1\to 2}(t_1^{\eta_3}) \mathcal{T}_{1\to 2}(t_1'^{\eta_4}) \rangle]  \label{Keldysh}
\end{align}
upto the leading order $O(\gamma_1^2 \gamma_2^2)$ of the perturbation,
where $\eta_i$'s are Keldysh indices, $T_\textrm{K}$ means the Keldysh ordering, and $t_2 = 0$ is chosen.
%{\color{blue}$W_{3 \to 2}$ has the same expression as Eq.~\eqref{Keldysh} but with the integrand replaced by $ \langle T_\textrm{K} \mathcal{T}^\dagger_{2\to 3}( t_2^{\eta_1=1}) \mathcal{T}_{2\to 3}(t_2'^{\eta_2=-1}) \mathcal{T}^\dagger_{1\to 2}(t_1^{\eta_3}) \mathcal{T}_{1\to 2}(t_1'^{\eta_4}) \rangle$.}

We compute $W_{i \to j}$ analytically to get Eq. (2) of the main text and also numerically. 
The analytic calculation can be done~\cite{Han,Kane_PRB} at $e^* V \gg k_B T$ where $W_{i \to j}$ is determined by the narrow domain of $|t_1 - t_1'| \lesssim \hbar / (e^*V)$, as $t_1$ ($t_1'$) is the time of injection of a particle-like anyon from Edge1 to Edge2 in subprocess $a_1$ ($a_2$).
The Poisson process for $W_{2 \to 3}^\textrm{P}$ occurs in the domain $t_2 \simeq t_2' \simeq t_1+d/v \simeq t_1'+d/v$ of the terms $\eta_3 = - \eta_4 = -1$ of $W_{2 \to 3}$ in Eq.~\eqref{Keldysh}.
The TVB (its partner disconnected process) for $W_{3 \to 2}^\textrm{TVB}$ occurs in the domain $t_2' <  t_1+d/v \simeq t_1'+d/v < t_2$ and $t_2< t_1+d/v \simeq t_1' + d/v < t_2'$ of the terms $\eta_3 = - \eta_4 = -1$  ($\eta_3 = \eta_4 = \pm 1$) of $W_{3 \to 2}$. The first (second) term of Eq. (5) in the main text is found from a term with  $\eta_3 = - \eta_4 = -1$ ($\eta_3 = \eta_4 = 1$) at $t_1' \to t_1$. The TVB and its partner for $W_{2 \to 3}^\textrm{TVB}$ similarly occur in $W_{2 \to 3}$.

\subsection{Leading-order correction to Eqs.~(1) and (2)}

We obtain the corrections of the leading order $O(T/V)$ to Eqs.~(1) and (2) below. 
At $e^*V \gg k_B T$, the corrections to the tunneling current $I$ and the excess noise $\delta S$ are found as
\begin{equation}
I \simeq e^* \gamma_1^2 \gamma_2^2 f(\nu) [\cos (\pi\nu)-\cos(3\pi\nu)]V^{2\nu-1} T^{2\nu-2} \left[ 1- 2\nu(1-2\nu)\tan(\pi\nu) \frac{\pi k_B T}{e^*V} \right],
\end{equation}
\begin{equation}
\delta S \simeq - 2{e^*}^2 \gamma_1^2 \gamma_2^2 f(\nu) [\cos(\pi\nu)-\cos(3\pi\nu)]V^{2\nu-1} T^{2\nu-2} \left[ 1+ 2\nu(1-2\nu)\cot(\pi\nu) \frac{\pi k_B T}{e^*V} \right].
\end{equation}
Combining the results, we find the leading-order correction to Eq. (1),
\begin{equation}
\frac{\delta S}{I} = - 2 e^*  \left[ 1 + \frac{4 \nu (1-2\nu)}{\sin 2 \pi \nu} \frac{\pi k_B T}{e^* V} \right].
\end{equation}

%We obtain the corrections of the leading order $O(T/V)$ to Eqs.~(1) and (2) below.  At $e^*V \gg k_B T$, the corrections to the tunneling current $I$ and the excess noise $\delta S$ are found as
%\begin{equation}
%I \simeq e^* \frac{\gamma_1^2\gamma_2^2  \tau_0^{4\nu} }{\left(2\pi a\right)^4} \frac{\pi 2^{2\nu+1}\cos(\pi\nu) \left( 1-\cos(2\pi\nu) \right)}{\nu^2 \Gamma(2\nu)}  \left(\frac{e^*V}{\hbar} \right)^{2\nu-1} (\pi T)^{2\nu-2} \left[ 1- 2\nu(1-2\nu)\tan(\pi\nu) \left( \frac{\pi T}{e^*V} \right) \right],
%\end{equation}
%\begin{equation}
%\delta S \simeq-2\left(e^*\right)^2 \frac{\gamma_1^2\gamma_2^2  \tau_0^{4\nu} }{\left(2\pi a\right)^4} \frac{\pi 2^{2\nu+1}\cos(\pi\nu) \left( 1-\cos(2\pi\nu) \right)}{\nu^2 \Gamma(2\nu)}  \left(\frac{e^*V}{\hbar} \right)^{2\nu-1} (\pi T)^{2\nu-2} \left[ 1+ 2\nu(1-2\nu)\cot(\pi\nu)\left(\frac{\pi T}{e^*V} \right) \right]
%\end{equation}
 
%Negative excess noise $\delta S=-2e^*I$ indicates presence of fractional statistics but cannot specify the fractional parameter $\nu$ corresponding to particular fractional statistics. Therefore we show here that fractional parameter can be extracted from the voltage-dependence of the ratio, $\frac{\delta S}{I}$ at fixed temperature. The result $\delta S=-2e^*I$ is obtained by approximating the voltage biased anyon, whose wavepacket size is $\frac{\hbar v_P}{e^*V}$, with a particle localized at a point. Therefore the correction terms are $\left(\frac{1}{e^*V}\right)^n$s.   By taking only term of $\left(\frac{1}{e^*V}\right)$ order, we obtain in the shot noise regime ($e^*V \gg k_B T$), 

 \begin{figure}[b] 
\centering
\includegraphics[width=.95\columnwidth]{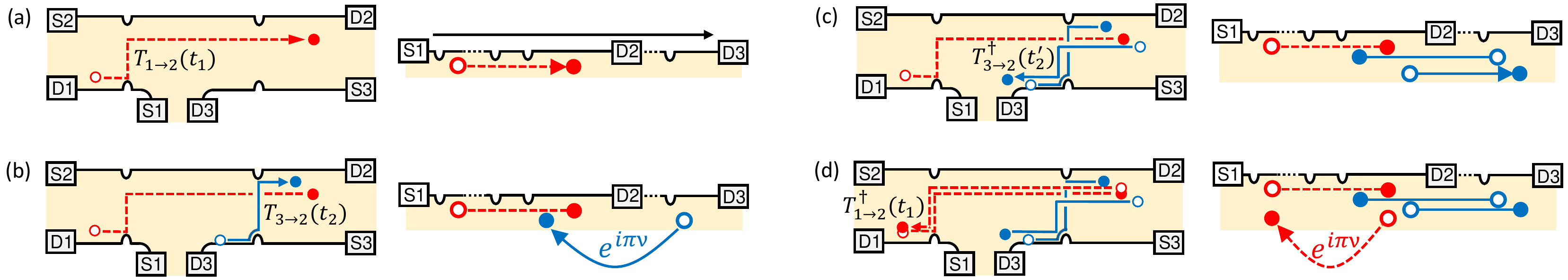}
\caption{Anyon propagations in the subprocesses [the left panels of (a-d)] of the interference $a^*_2 a_1$  are drawn on the connected edge [the right panels] corresponding to our setup. We used the connected-edge scheme in Fig.~\ref{connected_edge}. A voltage-biased particle-like (hole-like) anyon is depicted by the filled (empty) red circles. A thermal particle-like (hole-like) anyon is drawn as the filled (empty) blue circles.}  \label{effective_braiding} 
\end{figure}

\subsection{Effective braiding}

The effective braiding phase factor $e^{2i\pi\nu}$ in Eq.~(6) can be understood by using the connected-edge scheme in Fig.~\ref{connected_edge}. One half $e^{i\pi\nu}$ of the phase factor is gained in the subprocess $a_1$. Here, a pair of a voltage-biased particle-like anyon and its hole-like partner is excited at QPC1, and then the anyons propagate. This process, described by the tunneling operator $T_{1 \to 2} (t_1)$, is drawn on the connected edge in the right panel of Fig.~\ref{effective_braiding}(a). After the voltage-biased particle-like anyon passes QPC2, a pair of thermal particle-like and hole-like anyons is excited at QPC2. This is described by $T_{3 \to 2} (t_2)$. The resulting configuration of the four anyons by $T_{3 \to 2} (t_2) T_{1 \to 2} (t_1)$ is shown in  Fig.~\ref{effective_braiding}(b). Notice that the thermal particle-like anyon is in the left side of the voltage-biased particle-like anyon on the connected edge, while the thermal hole-like anyon is created in the right side. This is equivalent with the following situation: The thermal particle-like and hole-like anyons are created from the vacuum on the right side of the voltage-biased particle-like anyon, and then the thermal particle-like anyon moves to the left side. In this anyon exchange,  a phase factor $e^{i\pi\nu}$ is gained. This is the one half of the braiding phase factor, and obtained partly by a Klein factor in the computation. 

On the other hand, the other half  $e^{i\pi\nu}$ of the braiding phase factor is gained in interference $a^*_2 a_1$ between $a_1$ and $a_2$. In $a_2$, a pair of a voltage-biased particle-like anyon and its hole-like partner is excited at QPC1, and then the anyons propagate. This is described by $T_{1 \to 2} (t_1)$.  Before the voltage-biased particle-like anyon passes QPC2, a pair of thermal particle-like and hole-like anyons is excited at QPC2. This is described by $T_{3 \to 2} (t_2')$. In the interference $a^*_2 a_1$ described by $T^\dagger_{1 \to 2} (t_1) T^\dagger_{3 \to 2} (t_2') T_{3 \to 2} (t_2) T_{1 \to 2} (t_1)$, we consider the complex-conjugated process of $a_2$, $T^\dagger_{1 \to 2} (t_1) T^\dagger_{3 \to 2} (t_2')$. On top of the anyon configuration of the subprocess $a_1$ [$T_{3 \to 2} (t_2) T_{1 \to 2} (t_1)$; see Fig.~\ref{effective_braiding}(b)], we first draw the anyon configuration corresponding the excitation of the thermal anyon pair of $a^*_2$ on the connected edge [see Fig.~\ref{effective_braiding}(c)]. Here, the pair is created in the outside of the dashed line between the voltage-biased particle-like anyon and its hole-like partner. Therefore, no anyon exchange happens between the voltage-biased anyons and the thermal anyons in this step. We next draw the voltage-biased anyon pair of $a_2^*$ on the connected edge [see Fig.~\ref{effective_braiding}(d)]. We notice that the voltage-biaed paritcle-like anyon is created in the left side of the thermal particle-like anyon of $a_1$, while the voltage-biased hole-like anyon occurs in the right side. This is equivalent with the following situation: The voltage-biased particle-like and hole-like anyons of $a_2^*$ are created from the vacuum on the right side of the thermal particle-like anyon of $a_1$, and then the voltage-biased particle-like anyon moves to the left side. In this anyon exchange,  a phase factor $e^{i\pi\nu}$ is gained. This is the other half of the braiding phase factor.

%In this subsection, we concretize the concept of effective braiding and explain how the braiding phase factor $e^{2i\pi\nu}$ arises in Fig.~2 of the main text. `Effective braiding' consists of two position exchanges of anyons in the same direction in chiral 1D. A position exchange corresponds to an intersection point between two lines in Fig.~2, which is easily seen by comparing $a_1$ of Fig.~2 (a) with (2) of Fig.~\ref{effective_braiding}. The braiding phase factor $e^{2i\pi\nu}$ results from two intersection points between two red lines and a blue line in Fig.~2 (b). Two position exchanges corresponding to the two intersection points are in the same direction and therefore their phases add up to braiding phase factor $e^{2i\pi\nu}$. Opposite direction of a red line comparing to the other one reverse the direction of exchange, but oppoiste packing order(the red on top of the blue and the blue on top of the red) once again reverse the direction of exchange.
 
\subsection{Effects of an additional co-propagating (downstream) outer edge channel}

We consider the situation where there is an additional co-propagating (downstream) outer channel along Edge1, Edge2, and Edge3. This situation can occur in fractional quantum Hall systems of, e.g., filling factor $7/3 = 2 + 1/3$ or $4/3 = 1 + 1/3$~\cite{An,Baer}. In this situation, the negative excess noise $\delta S < 0$ is still found. The value of $\delta S$ depends on the inter-channel interaction strength and the distance between QPC1 and QPC2 in Edge2. The negative excess noise provides a signature of the fractional statistics of anyons in the inner channel. We expect that the outer channel is helpful for observing the negative excess noise, since it can screen the effect~\cite{Rosenow} of edge reconstruction or upstream neutral modes; the effect of the neutral modes will be discussed in the next section.
  
The Hamiltonian in Eq.~\eqref{totH_o} is modified as $H = \sum_i H_{i} + H_{i,\textrm{out}} + H_\textrm{tun}$ to include the additional outer edge channel,
%Edge$i$ is described by the chiral Luttinger liquid~\cite{Wen,vonDelft} \begin{equation} H_{i} =\frac{\hbar v}{4\pi \nu} \int_{-\infty}^{\infty} dx :(\partial_x\phi_i(x))^2:+e^* V_i \hat{N}_i, \end{equation} the co-propagating phonon mode and its coupling to Edge$i$ is described by
 \begin{equation} 
H_{i,\textrm{out}}=\frac{\hbar v_\textrm{out}}{4\pi \nu_\textrm{out}} \int_{-\infty}^{\infty} dx :(\partial_x\phi_{i,\textrm{out}}(x))^2:+\frac{\hbar \kappa}{2\pi} \int_{-\infty}^{\infty} dx \partial_x\phi_i(x) \partial_x\phi_{i,\textrm{out}}(x). \label{Hiout} \end{equation} 
The first term is the Hamiltonian of the outer edge channel, and the second term describes the interaction between the density $\partial_x\phi_i(x)$ of the inner edge channel and the density $\partial_x\phi_{i,\textrm{out}}(x)$ of the outer channel. $v_\textrm{out}$ is the velocity of the outer channel, $\nu_\textrm{out}$ is the effective filing factor for the outer channel, and $\kappa$ is the inter-channel interaction strength. The commutation relations of the boson fields are \begin{equation}
[\phi_i(x),\phi_i(x')]=i\pi\nu \textrm{sign}(x-x'), \quad \quad [\phi_{i,\textrm{out}}(x),\phi_{i,\textrm{out}}(x')]=i\pi\nu_{\textrm{out}} \textrm{sign}(x-x'), \quad \quad [\phi_i(x),\phi_{i,\textrm{out}}(x')]=0.
\end{equation}
As in the case of the single edge channel per edge, anyon tunneling occurs between the inner channels (described by $\phi_i(x)$'s) of Edge1 and Edge2 at QPC1, and between the inner channels of Edge2 and Edge3 at QPC2, and we treat the tunneling Hamiltonian $H_\textrm{tun}$ as a perturbation on $\sum_i H_i+H_{i,\textrm{out}}$.

The bare Hamiltonian $\sum_i H_i+H_{i,\textrm{out}}$ is diagonalized as
\begin{equation} H_i+H_{i,\textrm{out}}=\frac{\hbar v_1}{4\pi \nu_1} \int_{-\infty}^{\infty} dx :(\partial_x\phi_{i1}(x))^2: + \frac{\hbar v_2}{4\pi \nu_2} \int_{-\infty}^{\infty} dx :(\partial_x\phi_{i2}(x))^2: + e^* V_i \hat{N}_i
\end{equation}
\begin{equation}\begin{pmatrix}\phi_{i}(x) \\  \phi_{i,\textrm{out}}(x)
\end{pmatrix} = \begin{pmatrix} T_{11} & T_{12} \\ T_{21} & T_{22} 
\end{pmatrix}   \begin{pmatrix} \phi_{i1}(x) \\  \phi_{i2}(x)  
\end{pmatrix}  \end{equation}
where the commutation relations for the new boson fields are \begin{equation}
[\phi_{i1}(x),\phi_{i1}(x')]=i\pi\nu_1 \textrm{sign}(x-x'), \quad \quad [\phi_{i2}(x),\phi_{i2}(x')]=i\pi\nu_{2} \textrm{sign}(x-x'), \quad \quad [\phi_{i1}(x),\phi_{i2}(x')]=0.
\end{equation}
The elements of the transformation matrix $T$ are found as
%By comparing cross term($\phi_{i1}(x) \phi_{i2}(x)$) in hamiltonian and three commutation relations, 
%\begin{align}
%\quad &\nu=\nu_1 T_{11}^2 + \nu_2 T_{12}^2, \quad \nu_P = \nu_1 T_{21}^2 + \nu_2 T_{22}^2, \quad \nu_1T_{11}T_{21}+\nu_2T_{12}T_{22}=0, \nonumber \\
%&\frac{v}{\nu}T_{11}T_{12}+\frac{v_P}{\nu_P}T_{21}T_{22}+\kappa \left( T_{11}T_{22}+T_{12}T_{21} \right)=0
%\end{align} 
%which determine $\begin{pmatrix} T_{11} & T_{12} \\ T_{21} & T_{22} 
%\end{pmatrix}$. Then the diagonalized veolocities are \begin{equation}
%\frac{v_1}{\nu_1}=\frac{v}{\nu} T_{11}^2 + \frac{v_P}{\nu_P} T_{21}^2 + 2\kappa T_{11}T_{21}, \quad \quad \frac{v_2}{\nu_2}=\frac{v}{\nu} T_{12}^2 + \frac{v_P}{\nu_P} T_{22}^2 + 2\kappa T_{12}T_{22}
%\end{equation}
%We can set $\nu_1=1$ and $\nu_2=1$.
\begin{align}
&T_{11}=\left[ \frac{\nu}{2}\left(1+ \frac{(v-v_\textrm{out})}{\sqrt{(v-v_\textrm{out})^2+4\kappa^2\nu\nu_\textrm{out}}} \right) \right]^{1/2} , \quad \quad T_{12}=\left[ \frac{\nu}{2}\left(1- \frac{(v-v_\textrm{out})}{\sqrt{(v-v_\textrm{out})^2+4\kappa^2\nu\nu_\textrm{out}}} \right) \right]^{1/2} \\ \nonumber &
T_{21}=\left[ \frac{\nu_\textrm{out}}{2}\left(1- \frac{(v-v_\textrm{out})}{\sqrt{(v-v_\textrm{out})^2+4\kappa^2\nu\nu_\textrm{out}}} \right) \right]^{1/2}, \quad  T_{22}=-\left[ \frac{\nu_\textrm{out}}{2}\left(1+ \frac{(v-v_\textrm{out})}{\sqrt{(v-v_\textrm{out})^2+4\kappa^2\nu\nu_\textrm{out}}} \right) \right]^{1/2}.
\end{align}
Using the diagonalized form, we obtain
the correlator $\left\langle \phi_i(0,0) \phi_i(x,t)  \right\rangle_0$ useful for computing the current $I$ and the excess noise $\delta S$,
\begin{align}
&\left\langle \phi_i(0,0) \phi_i(x,t)  \right\rangle_0 = \left\langle \left( T_{11} \phi_{i1}(0,0)+T_{12} \phi_{i2}(0,0) \right) \left( T_{11} \phi_{i1}(x,t)+T_{12} \phi_{i2}(x,t) \right)  \right\rangle_0 \\ \nonumber &= T_{11}^2 \left\langle \phi_{i1}(0,0)\phi_{i1}(x,t)  \right\rangle_0 + T_{12}^2 \left\langle \phi_{i2}(0,0)\phi_{i2}(x,t)  \right\rangle_0= \textrm{ln}\left[ \left(\frac{\tau_0+i(t-x/v_1)}{\tau_0} \right)^{T_{11}^2} \left(\frac{\tau_0+i(t-x/v_2)}{\tau_0} \right)^{T_{12}^2}  \right],
\end{align}
where
\begin{equation}
v_1=\frac{v+v_\textrm{out}+\sqrt{(v-v_\textrm{out})^2+4\kappa^2\nu\nu_\textrm{out}}}{2}, \quad \quad  v_2=\frac{v+v_P-\sqrt{(v-v_\textrm{out})^2+4\kappa^2\nu\nu_\textrm{out}}}{2}.
\end{equation}

%\begin{align}
%&I=e^*(W_{2 \rightarrow 3} - W_{3 \rightarrow 2}) \\ \nonumber
%&\delta S = 2 \left(e^* \right)^2 (W_{2 \rightarrow 3} + W_{3 \rightarrow 2}) 
%\end{align}
%where $W_{2\rightarrow 3}$ and $W_{3 \rightarrow 2}$ are expressed in S2. The tunnelig operators in $W_{i \rightarrow j}$ are $\mathcal{T}_{1\rightarrow 2}(t_1) = \Psi_2^{\dagger}(0,t_1)\Psi_1(0,t_1)$ and  $\mathcal{T}_{3 \to 2}(t_2) = \Psi_2^\dagger (d,t_2) \Psi_3 (0,t_2)$ where $\Psi^\dagger_i (x,t) = F_i^\dagger (t) e^{-i \phi_i (x,t)} / \sqrt{2 \pi a}$.

%By using identity\cite{vonDelft} $\langle e^{s_1 \phi_1} e^{s_2 \phi_2} \cdots \rangle = e^{\sum_j s^2_j \langle \phi_j^2 \rangle + \sum_{i<j} s_i s_j \langle \phi_i \phi_j \rangle}$ with $s_i = \pm i$ and the correlator $\left\langle \phi_i(0,0) \phi_i(x,t)  \right\rangle_0$, 

\begin{figure}[b] 
\centering
\includegraphics[width=0.9\columnwidth]{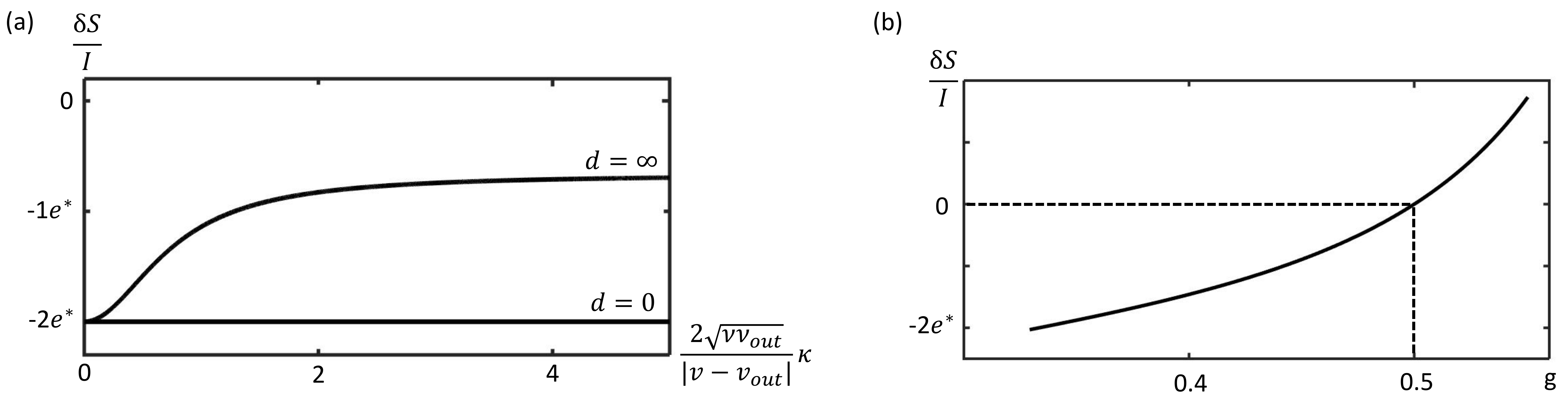}
\caption{(a) Current-to-noise ratio $\delta S / I$ as a function of the normalized interaction strength $\frac{2\sqrt{\nu \nu_\textrm{out}}}{\mid v-v_\textrm{out} \mid}\kappa$ in the case of an additional downstream (co-propagating) edge channel. $\nu = 1/3$ is chosen, while $\nu_\textrm{out}$ is arbitrary. (b) $\delta S / I$ as a function of the exponent $g$ of the current-voltage characteristics of anyon tunneling at the QPCs in the presence of edge reconstruction. $\nu = 1/3$ is chosen.}  \label{ratio_kappa_coprop} 
\end{figure}

Using the correlator, we obtain $\delta S / I$ at $e^*V \gg k_BT$,
\begin{equation}
\frac{\delta S}{I}=-2e^* \cot{(\pi \nu)} \frac{[1-\cos(2\pi\nu)]e^{-Ad}+ [2-\cos(2\pi T_{11}^2)-\cos(2\pi T_{12}^2)](1-e^{-Ad}) }{\sin(2\pi\nu) e^{-Ad}+ [\sin(2\pi T_{11}^2)+\sin(2\pi T_{12}^2)] (1-e^{-Ad})},
\end{equation}
where $e^* = \nu e$, $A=2 \nu \left( \frac{1}{v_2}-\frac{1}{v_1} \right) \pi T$, and  $d$ is the distance between QPC1 and QPC2 on Edge2. When $d=0$, we find $\delta S /I =-2e^*$, which is independent of the inter-channel interaction strength $\kappa$ and the properties (such as $v_\textrm{out}$ and $\nu_\textrm{out}$) of the outer edge channel.
On the other hand, in the case of $d \to \infty$ (much longer than the spatial broadening of thermal anyons excited at QPC2), $\delta S / I$ increases with increasing the interaction strength $\kappa$.
At $\kappa \to \infty$ and $d \to \infty$, $\delta S / I$ converges to
\begin{equation} \frac{\delta S}{I} \to - 2 e^* \cot (\pi \nu) \frac{1 - \cos (\pi \nu)}{\sin ( \pi \nu)}. \end{equation}
For example, at $\nu = 1/3$ (i.e., at the filling factor $4/3 = 1 + 1/3$ or $7/3 = 2 + 1/3$), $\delta S / I = - 2e^* / 3$ at $\kappa \to \infty$ and $d \to \infty$.  See Fig.~\ref{ratio_kappa_coprop}(a).
This shows that the excess noise is always negative even in the presence of additional co-propagating (downstream) channels along the edges.  
 
\subsection{Effects of edge reconstruction}
 
In this section, we study the effect of edge reconstruction. The edge reconstruction can result in upstream outer channels~\cite{Rosenow}. To see the effect, we consider the situation where there is an additional upstream outer channel along each edge (instead of an additional downstrem channel as in the last section). In this situation, the excess noise $\delta S$ is negative only when the interchannel interaction strength is sufficiently weak. We also discuss the possibility of observing the negative excess noise $\delta S<0$ in the situation of some reported experiments.

The Hamiltonian in Eq.~\eqref{totH_o} is now modified as $H = \sum_i H_{i} + H_{i,\textrm{P}} + H_\textrm{tun}$. The term $H_{i,\textrm{P}}$ describes the upstream outer channel (which is a phonon mode in Ref.~\cite{Rosenow}) and its interaction with the downstream inner channel, 
\begin{equation} 
H_{i, \textrm{P}}=\frac{\hbar v_\textrm{P}}{4\pi \nu_\textrm{P}} \int_{-\infty}^{\infty} dx :(\partial_x\phi_{i, \textrm{P}}(x))^2:+\frac{\hbar \kappa}{2\pi} \int_{-\infty}^{\infty} dx \partial_x\phi_i(x) \partial_x\phi_{i, \textrm{P}}(x). \end{equation} 
It has the same form as $H_{i,\textrm{out}}$ in Eq.~\eqref{Hiout}. The upstream outer channel is described by the boson field $\phi_{i,\textrm{P}}$. 
The commutation relations of the boson fields are \begin{equation}
[\phi_i(x),\phi_i(x')]=i\pi\nu \textrm{sign}(x-x') \quad \quad [\phi_{i, \textrm{P}}(x),\phi_{i, \textrm{P}}(x')]=-i\pi\nu_\textrm{P} \textrm{sign}(x-x') \quad \quad [\phi_i(x),\phi_{i, \textrm{P}}(x')]=0
\end{equation}
The minus sign in the second commutation relation implies the upstream flow of the outer channel.
 
Following the steps discussed in the last section, we obtain
\begin{align}
\frac{\delta S}{I}=-2e^* \frac{\tan \frac{\pi (\nu+g)}{2}}{\tan\left(\pi g \right)} \label{edgereconst}
\end{align}
for $g < 1$.
Here $g = \frac{ \nu   (v+v_P)}{\sqrt{(v+v_P)^2-4\kappa^2\nu\nu_P}}$ is the exponent of the relation~\cite{Wen} $\mathcal{I} \sim \mathcal{V}^{2g-1}$ of anyon tunneling current $\mathcal{I}$ between two edges at a QPC by a bias voltage $\mathcal{V}$.
In the absence of the edge reconstruction, $g$ is identical to the filling factor $\nu$ (in the case of the Laughlin-sequence filling factors), and the above relation becomes the expression $\delta S / I = - 2 e^*$ in Eq. (1). In the presence of edge reconstruction, $g$ becomes larger than $\nu$.
In Eq.~\eqref{edgereconst}, we find the expression of $\delta S / I$ as a function of $g$ (rather than as a function of $\kappa$) since $g$ can be measured in experiments.

As shown in Eq.~\eqref{edgereconst}, the excess noise $\delta S$ is negative when $g < 1/2$. The negative excess noise is a signature of the fractional statistics of anyons propagating in the inner channels. In the opposite case of $g > 1/2$, $\delta S$ is positive. For $\nu=1/3$, the dependence of $\delta S$ on $g$ is drawn in Fig.~\ref{ratio_kappa_coprop}(b). Note that the results are independent of the distance $d$ between QPC1 and QPC2 on Edge2, unlike the case of an additional co-propagating outer edge channel discussed in the previous section. The independence is due to the fractionalization of a voltage-biased anyon into two counter-propagating parts on Edge2.
 
There have been experimental reports~\cite{Saminadayar,Roddaro} of observing $g < 1/2$ in a FQH system with filling factor 1/3. The fact that $g$ is observed to be larger than $1/3$ implies that there occurs edge reconstruction.
In the cases of the reports of observing $g < 1/2$, we expect the possiblity that one can observe the negative excess noise $\delta S <0$, a signature of the fractional statistics.

 % The anyon tunneled through QPC1 into Edge2 is divided into a mode that copropagates with the original mode and a counterpropagating mode. Since the counterpropagating mode does not arrive at QPC2, the time between the arrivals($\infty$) of two mode is independent of $d$. The ratio as a function of the tunnelig exponent is in Fig. \ref{ratio_tun_expo_counterprop}.  
%\begin{align}
%\frac{\delta S}{I}=-2e^* \frac{\tan\left(\pi (\nu'+\nu)/2 \right)}{\tan\left(\pi \nu' \right)} \left[1+ \frac{2\left( 1-2\nu' \right)}{ \sin\left(\pi (\nu'+\nu)/2 \right)\cos\left(\pi (\nu'+\nu)/2  \right) } \left( \frac{\nu'+\nu}{2}+\frac{\nu'-\nu}{2} e^{-2\nu'\left( \frac{1}{v_1}+\frac{1}{v_2} \right)d} \right) \left(\frac{\pi T}{e^*V}\right) \right ]
%\end{align}